\documentclass{amia}
\usepackage{graphicx}
\usepackage[labelfont=bf]{caption}
\usepackage[superscript,nomove]{cite}
\usepackage{color}
\usepackage{hyperref}
\usepackage{color,hyperref}
\definecolor{darkblue}{rgb}{0.0,0.0,0.3}
\hypersetup{colorlinks,breaklinks,
            linkcolor=darkblue,urlcolor=darkblue,
            anchorcolor=darkblue,citecolor=darkblue}

\begin{document}

\title{Accuracy of the Epic Sepsis Prediction Model in a Regional Health System}

\author{Tellen D. Bennett, MD, MS$^{1,2}$, Seth Russell, MS$^{2}$, James King, MIDS$^{2}$,
Lisa Schilling, MD,MSPH$^{2,3}$, Chan Voong, MUSA$^{2}$, Nancy Rogers, BA,PMP$^{4}$, Bonnie Adrian, PhD,RN$^{4,5}$,
Nicholas Bruce, PhD$^{6}$, Debashis Ghosh, PhD$^{2,7}$}
\bigskip
\institutes{
    $^1$Pediatric Critical Care, University of Colorado School of Medicine, Aurora, CO; 
    $^2$CU Data Science to Patient Value (D2V), Anschutz Medical Campus, Aurora, CO;
    $^3$General Internal Medicine, University of Colorado School of Medicine, Aurora, CO;
    $^4$Clinical Informatics, University of Colorado Health, Aurora, CO;
    $^5$University of Colorado College of Nursing, Aurora, CO;
    $^6$Epic Corporation, Verona, WI;
    $^7$Biostatistics and Informatics, Colorado School of Public Health, Aurora, CO;
    \\
}

\maketitle

% \noindent{\bf Abstract}
% 
% \textit{An accurate electronic health record-based predictive model might allow early treatment of sepsis, one of the leading causes of morbidity and mortality worldwide. We evaluated the accuracy of the proprietary Epic sepsis prediction model (ESPM) during 8,206 adult admissions in a five-hospital regional health system. The ESPM predicted sepsis with a positive predictive value of 0.44, recall of 0.66, negative predictive value of 0.91, F1-measure of 0.53, and area under the curve of 0.73.}

\section*{Introduction}
Sepsis, "life-threatening organ dysfunction caused by a dysregulated host response to infection,"\cite{Singer2016s3} is an important public health problem and one of the leading causes of morbidity and mortality worldwide. Sepsis is difficult to detect and diagnose because the pathobiology is not completely understood and no validated criterion standard diagnostic test exists. Because some\cite{Rivers2001} data suggest that early aggressive treatment of sepsis improves outcomes, many health systems attempt to rapidly identify patients with sepsis. Our institution, University of Colorado Health (UCHealth), has developed a sepsis detection program for this purpose based on the modified Early Warning Score (EWS)\cite{Churpek2017}.

Interest in an electronic health record (EHR)-based computational model that can accurately predict a patient's risk of sepsis at a given point in time has grown rapidly in the last several years.\cite{Desautels2016} Like other EHR vendors\cite{Amland2016}, the Epic Corporation has developed a proprietary sepsis prediction model (ESPM). Epic developed the model using data from three health systems and penalized logistic regression. Demographic, comorbidity, vital sign, laboratory, medication, and procedural variables contribute to the model. UCHealth obtained a license for that model in 2017. The objective of this project was to compare the predictive performance of the ESPM with UCHealth's current EWS-based program. 

\section*{Methods}

\textit{Design}. Retrospective cohort study. University institutional review board approval was obtained.

\textit{Setting/Patients}. All inpatients at five hospitals in a regional health system between June 1, 2017 and October 20, 2017 who at any time during their hospitalization were admitted to one of the 15 units (three at each hospital) where the ESPM was calculated. Eleven of the units were inpatient units and four were intensive care units.

\textit{Primary Predictors/Variable Definitions}. The primary predictors were the sequentially updated ESPM and EWS. The ESPM was automatically calculated approximately every two hours for all inpatients on the 15 UCHealth units. The output is a whole number between 0 and 100 that represents the model-estimated probability of developing sepsis given the information available at that time. The EWS, a whole number between 0 and 21, is automatically calculated every four hours and then manually filed by bedside nurses. We defined comorbidities such as chronic conditions and oncology status using freely available software (\url{https://www.hcup-us.ahrq.gov/toolssoftware/ccs10/ccs10.jsp}).

\textit{Main Outcome}. The primary outcome was the presence of any International Classification of Diseases, 10th Revision (ICD-10) diagnosis code for sepsis, as defined in a proprietary multi-institutional benchmarking codeset (Vizient).

\section*{Results}

\textit{Patient Characteristics and Hospital Outcomes}. Overall, 7,427 patients met inclusion criteria. They experienced 8,206 hospitalizations during the study period. The cohort was typical of adult hospital inpatients: the mean age was 56.5 years and nearly all (7,748/8,206, 94\%) of the patients had at least one chronic condition. Sepsis occurred in 1,507/8,206 (18\%) of hospitalizations. Patients who developed sepsis tended to be older (mean age 59.5 versus 55.9 years, t-test \textit{P} $<$ 0.001) and were more likely to have oncologic conditions (20\% versus 13\%, chi-square \textit{P} $<$ 0.001). Mortality was much more common in patients with sepsis than those without sepsis (10\% versus 1\%, chi-square \textit{P} $<$ 0.001). The median time to in-hospital death was 5 days (interquartile range [IQR] 2 to 10). 

\textit{ESPM and EWS Performance}. The first values for each patient occurred a median of 7 hours (ESPM, IQR 4 to 12) versus 10 hours (EWS, IQR 6 to 32) after admission. Nearly all (97\%) of the patients had at least one ESPM and at least one EWS value $>$ 0. The maximum values were 66 (ESPM) versus 15 (EWS) among patients with sepsis and 40 (ESPM) versus 16 (EWS) among patients without sepsis. Patients tended to have their highest ESPM values earlier in the hospital course (Figure 1a). Patients with sepsis overall had higher ESPM values than those without sepsis. Using a threshold value of 5 for both scores (Figure 1b), the ESPM predicted sepsis better than the EWS: positive predictive value of 0.44 versus 0.33, recall of 0.66 versus 0.61, negative predictive value of 0.91 versus 0.85, F1-measure of 0.53 versus 0.43, area under the receiver operator characteristic curve (AUC) of 0.73 versus 0.62, and accuracy of 0.78 versus 0.64. However, the ESPM identified patients after some may have already come to the attention of UCHealth's existing sepsis detection program: it reached the threshold of 5 a median of 7 hours after the first lactate level was drawn (IQR 4 to 22).

% More text of an additional paragraph, with a figure reference (Figure ~\ref{fig1}) and a figure inside a Word text box below.  Figures need to be placed as close to the corresponding text as possible and not extend beyond one page.\\
\begin{figure}[h!]
\centering
\includegraphics{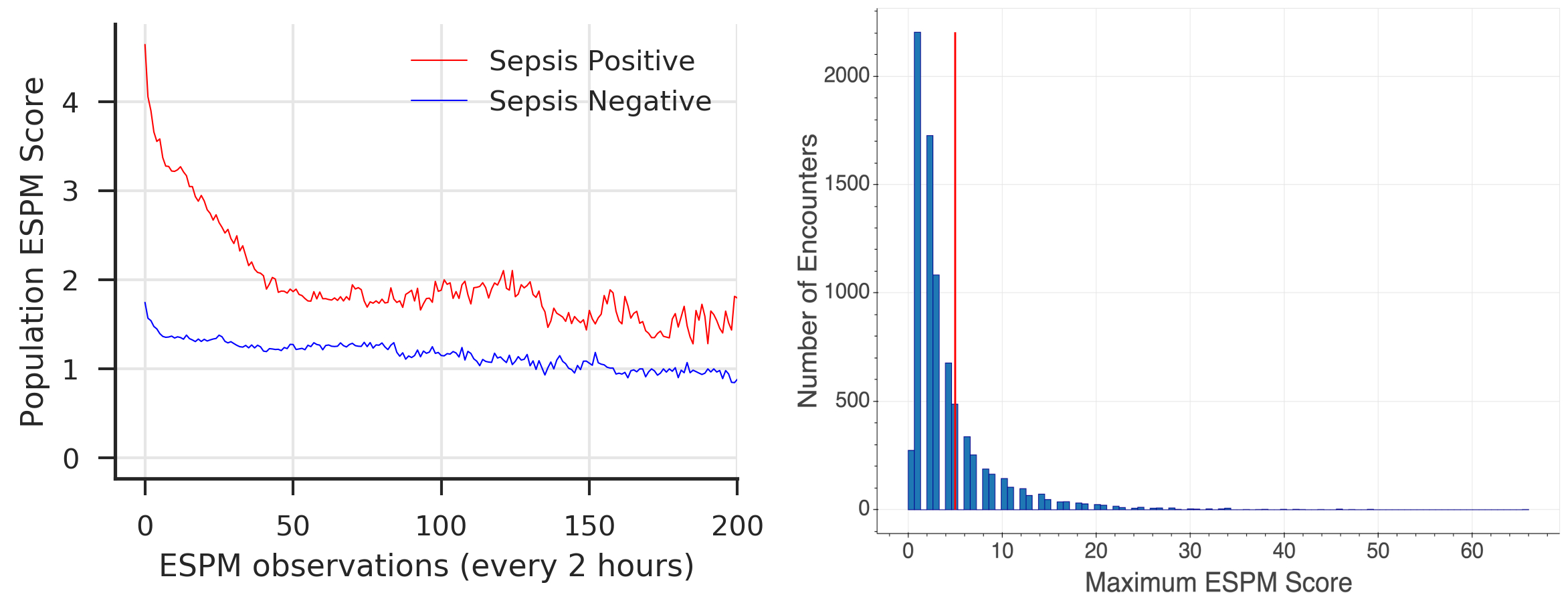}
\caption{Time series of average ESPM scores (1a) and Histogram of maximum ESPM score (1b) with threshold = 5}
\label{fig1}
\end{figure}

\section*{Discussion}
The ESPM predicts sepsis moderately accurately in a five-hospital regional health system. It is more accurate than the current EWS-based system. The ESPM does not currently run in Emergency Departments or on every inpatient unit in our health system, which may explain some of the differences in timing of sepsis detection.

\renewcommand{\refname}{References} % Rename the bibliography section.

\end{document}